\documentclass[traditabstract,print]{aa}
\usepackage{graphicx}
\usepackage{epsfig}
\usepackage{amssymb,amsmath}
\usepackage{natbib}
\bibpunct{(}{)}{;}{a}{}{,}
\usepackage{pslatex}
\usepackage{caption}
\usepackage{subfigure}
\usepackage{hyperref}
\usepackage{gensymb}
\usepackage{multirow}

\begin{document}

\title{Lensing of fast radio bursts by binaries to probe compact dark matter}

\author{Y. K. Wang$^{1}$ and F. Y. Wang$^{1,2}$\thanks{E-mail: \href{fayinwang@nju.edu.cn}{fayinwang@nju.edu.cn}}}

\institute{
$^{1}$ School of Astronomy and Space Science, Nanjing University, Nanjing 210093, China\\
$^{2}$ Key Laboratory of Modern Astronomy and Astrophysics (Nanjing
University), Ministry of Education, Nanjing 210093, China}

\authorrunning{Wang \& Wang}
\titlerunning{Lensing of FRBs to Probe Compact Dark Matter }

\abstract{The possibility that a fraction of dark matter is comprised of massive compact halo objects
(MACHOs) remains unclear, especially in the 20-100  $M_{\odot}$ window. MACHOs could
make up binaries, whose mergers may be detected by LIGO as
gravitational wave events. On the other hand, the cosmological origin of fast radio burst (FRBs) has been confirmed.
We investigate the possibility of
detecting FRBs gravitational lensed by MACHO
binaries to constrain their properties. Since lensing events could generate more than one image,
lensing by binaries could cause multiple-peak FRBs. The angular
separation between these images is roughly $10^{-3}$ mas, which is
too small to be resolved. The typical time interval between
different images is roughly 1 millisecond (ms). The flux ratio
between different images is from approximately 10 to $10^3$. With the
expected detection rate of $10^4$ FRBs per year by the upcoming
experiments, we could expect five multi-peak FRBs observed per year
with a time interval larger than 1 ms and flux ratio less than $10^3$
if the fraction of dark matter in MACHOs is $f\sim0.01$. A null
search  of multiple-peak FRBs for time intervals larger than 1 ms and
flux ratio less than $10^3$ with $10^4$ FRBs would constrain the
fraction $f$ of dark matter in MACHOs to $f<0.001$. }


\maketitle

\section{Introduction}
The nature of dark matter is an important question in cosmology.
Among many candidates, the weakly interacting massive particle (WIMP) is
a popular hypothesis but has not been proved. Alternatively, it has been proposed
that massive compact halo objects (MACHOs) could make up the dark
matter in our Universe
\citep{CarrHawking1974,Carr1975,Carr1976,Meszaros1975,Garcia-Bellido1996,Khlopov2010,Frampton2010,Belotsky2014}.
Many experiments have put constraints on the fraction of dark matter
that can reside in MACHOs with a given mass. Low-mass MACHOs ($<20
M_{\odot}$) are ruled out by microlensing surveys since they would
cause luminosity variability in stars
\citep{Alcock2001,Wyrzykowski2011,Tisserand2007}. High-mass MACHOs
($>100 M_{\odot}$) are excluded because they would disrupt wide
binaries \citep{Yoo2004,Quinn2009,Monroy2014}. The remaining
20-100 $M_{\odot}$ window is argued to be constrained by cosmic
microwave background (CMB), because the process of gas
accretion onto primordial black holes in the early universe could
modify the cosmic recombination history, which would affect the spectrum
and anisotropies of CMB \citep{Ricotti2007,Ricotti2008}. However, a
significant uncertainty is associated with this constraint
\citep{Ali2017} leaving the possibility that part of the dark matter could still be made of
MACHOs \citep{Pooley2009,Mediavilla2009}.

Recently, the advanced laser interferometer gravitational-wave observatory (LIGO) detectors recorded the first gravitational
wave (GW) event GW150914 \citep{Abbott2016a}, and the second GW
151226 \citep{Abbott2016b}. Both events are produced by the mergers
of black hole binaries, each with a mass of about $30M_\odot$.
However, in the traditional stellar evolution scenario, binaries of
black holes with masses of about $30M_\odot$ each are very difficult
to form. Some models have been proposed for the progenitors of
binary black hole systems
\citep{Hosokawa2016,Belczynski2016,Rodriguez2016a,Woosley2016,Chatterjee2017,Mink2016,Hartwig2016,Inayoshi2016,Arvanitaki2017,Rodriguez2016b}.
More interestingly, it has also been proposed that GW events could
possibly be the mergers of primordial black holes (PBHs)
\citep{Bird2016,Sasaki2016}, which could be treated as a part of
MACHOs.

Several ideas have been proposed to test the possibility that the
mergers of PBHs could be the progenitors of gravitational wave
events. It was suggested that the detectable residual eccentricity
when the two PBHs merge could help us test this possibility
\citep{Cholis2016}. If GW events are caused by the mergers of PBHs,
the GW events should have a low correlation with luminous galaxies.
The measurements of the cross-correlation of the GW events with
overlapping galaxy catalogs could help us test the model
\citep{Raccanelli2016}. The possibility of detecting stochastic GW
background from mergers of PBHs has been investigated
\citep{Mandic2016}. Strong gravitational lensing of extragalactic
fast radio bursts (FRBs) by PBHs is suggested to test the model
\citep{Munoz2016}.

FRBs are bright bursts of radio emission with a duration of a few milliseconds. The dispersion measures (DMs) of FRBs are much larger than
the line of sight DM contribution expected from the electron
distribution of our Galaxy. Recently, the precise localization of
the repeating FRB 121102 has revealed that it resides in a dwarf
galaxy at a redshift of $z=0.19$ \citep{Chatterjee2017,Tendulkar2017}, confirming its cosmological origin. FRBs are proposed to
measuring the cosmic proper distance \cite{Yu2017}. Many theoretical models for
FRBs are proposed, such as collapses of supra-massive neutron stars
\citep{Falcke14,Zhang14}, charged black hole binary mergers
\citep{Zhang2016}, collisions between asteroids and a high magnetized
pulsar \citep{Dai16}, superconducting cosmic strings \citep{Yu2014}
and double neutron star mergers \citep{Totani13,WangJ16}.

In this paper, we present the gravitational lensing properties of
fast radio bursts (FRBs) by PBH binaries and probe MACHOs.
FRBs are ideal sources to detect lensing events since they
are strong and short and occur at a rate about a few thousands per day
for the whole sky \citep{Lorimer07,Thornton13}. It has been
proposed that strong lensing of FRBs by isolated PBHs could be used
to probe MACHOs ($>20 M_{\odot}$) \citep{Munoz2016}. Considering
that the three GW events are produced by mergers of black hole
binaries, we study the strong lensing of FRBs by PBH binaries. In
this case, we could detect multiple-peak FRBs, unlike the repeating
FRBs. Since FRBs are strong, short bursts,
their multi-peak structure might be observable. In addition,
upcoming or ongoing surveys like APERTIF \citep{Verheijen2008},
UTMOST\citep{Caleb2016}, HIRAX \citep{Newburgh2016}, and CHIME
\citep{Bandura2014} expect to detect a large number of FRBs, providing a chance to detect multiple-peak FRBs. We adopted the
distribution of the separations of PBH binaries in
\cite{Nakamura1997} and \cite{Sasaki2016}, which assumes that PBHs'
velocities relative to the primordial gas have been redshifted to
negligible speeds when the gravitational attractions become
important. Without perturbation, two nearby black holes
would merge directly. However, the tidal force from other PBHs could
provide angular momentum for the two black holes to form a binary
system.

This paper is organized as follows. In Sect. \ref{sec:Optical
Depth}, we present the theories on strong lensing by binaries and give
our calculation on the optical depth for lensing by PBH binaries. In
Sect. \ref{sec:simulation}, we introduce our simulation of the
lensing events of FRBs by PBH binaries. Discussions and conclusion
are given in Sect. \ref{sec:conclusion}.

\section{Optical depth for lensing by PBH binaries}\label{sec:Optical Depth}

The theory of lensing by two-point-mass lenses \textbf{is} discussed
in detail by \cite{Schneider1986}. In the case of FRBs lensed by PBH
binaries, we treat two PBHs as two point lenses. $M=M_{1}+M_{2}$ is
defined as the total mass of the lenses, where $M_{1}$ and $M_{2}$
are the masses of two separate lenses. Throughout the paper, we consider the case $M_1=M_2= 30M_\odot$ as an
example. Then the Einstein radius is defined as
\begin{equation}\label{equation1}
\beta_{E}=\sqrt{\frac{4GM}{c^2}\frac{D_{d}D_{ds}}{D_{s}}},
\end{equation}
where $D_{d}$ is the angular diameter distance from the lens to the
observer, $D_{s}$ is the angular diameter distance from the source
to the observer, and $D_{ds}$ is the angular diameter distance
between lens and source. We define the following dimensionless
quantities
\begin{equation}\label{equation2}
\begin{aligned}
&\vec{r}=\vec{\xi}/\beta_E,\\
&\vec{x}=\left(\frac{D_{d}}{D_{s}}\right)\vec{\eta}/\beta_E,\\
&\mu_{1}=M_{1}/M,\\
&\mu_{2}=M_{2}/M,
\end{aligned}
\end{equation}
where $\vec{\eta}$ is the position of the source in the source plane,
and $\vec{\xi}$ is the position where light rays hit the lens plane.

Given the position of the source, there are between three and five image positions. The positions of images
in the lens plane are given by the lens equation \citep{Schneider1986},
\begin{equation}\label{equation6}
\vec{x}=\vec{r}-\vec{\alpha}\left(\vec{r}\right),
\end{equation}
where
\begin{equation}\label{equation7}
\vec{\alpha}\left(\vec{r}\right)=
\mu_{1}\frac{\vec{r}-\vec{r}_{0}}{|\vec{r}-\vec{r}_{0}|^{2}}+\mu_{2}\frac{\vec{r}+\vec{r}_{0}}{|\vec{r}+\vec{r}_{0}|^{2}}.
\end{equation}
Lenses are located at the points $\beta_E\vec{r}_{0}$ and
$-\beta_E\vec{r}_{0}$. $\vec{r}_{0}=(X,0)$, where $X\geq0$ is the position of the lens on the lens plane.

With the positions of lenses, source and images, we can calculate
the amplification factor $I_{0}$ for every image, that is, the
ratio of the flux density of one of the images caused by light
deflection and the image when the lenses were absent
\begin{equation}\label{equation8}
I_{0}=|\det~A|^{-1},
\end{equation}
where $A=\partial\vec{x}/\partial\vec{r},$ and
\begin{equation}\label{equation9}
\det~A=1-\left(\frac{\mu_{1}}{|\vec{r}-\vec{r}_{0}|^{2}}
+\frac{\mu_{2}}{|\vec{r}+\vec{r}_{0}|^{2}}\right)^{2}+\frac{16\mu_{1}\mu_{2}X^{2}r_{2}^{2}}{|\vec{r}-\vec{r}_{0}|^{4}|\vec{r}+\vec{r}_{0}|^{4}}.
\end{equation}
The gravitational lensing time delay between two images
$\vec{r_a},\vec{r_b}$ can be calculated from \citep{Schneider1986}
\begin{equation}\label{equation10}
c\Delta
t=\beta_{E}^{2}\frac{D_{s}}{D_{d}D_{ds}}(1+z_d)[\Theta(\vec{r_a},\vec{x})-\Theta(\vec{r_b},\vec{x})],
\end{equation}
where $\Psi(\vec{r})$ is the gravitational potential of the mass
distribution and
$\Theta(\vec{r},\vec{x})=(\vec{r}-\vec{x})^{2}/2-\Psi(\vec{r})$,
$D_d$ is the angular diameter distance from the observer to the
lens, $D_s$ is the angular diameter distance from the observer to
the source, and $D_{ds}$ is angular diameter distance from the lens to
the source.

\begin{table*}
\centering \caption{The integrated optical depth $\bar{\tau}$ for
different parameters for a FRB number density proportional to
SFH.\label{table1}}
\begin{tabular}{|c|c|c|c|c|c|c|c|}
\hline
$t_{f},I_{f}$ &$10^{-4}s,10^{3}$&$10^{-3}s,10^{3}$&$10^{-4}s,10^{2}$&$10^{-3}s,10^{2}$&$10^{-4}s,10^{1}$&$10^{-3}s,10^{1}$\\
\hline
$f=10^{-3}$& $1.57\times10^{-4}$ & $9.50\times10^{-5}$ & $2.39\times10^{-5}$ & $1.05\times10^{-5}$& $1.24\times10^{-6}$ & $4.81\times10^{-7}$\\
\hline
$f=10^{-2}$& $5.53\times10^{-3}$ & $3.19\times10^{-3}$ & $8.15\times10^{-4}$ & $3.42\times10^{-4}$ & $4.36\times10^{-5}$ & $1.67\times10^{-5}$ \\
\hline
$f=10^{-1}$& $5.74\times10^{-2}$ & $3.31\times10^{-2}$ & $8.44\times10^{-3}$ & $3.58\times10^{-3}$ &$4.62\times10^{-4}$ &$1.72\times10^{-4}$ \\
\hline
\end{tabular}
\end{table*}

In order to find all of the image positions for any given source
position, we follow the search routine for image positions proposed
in \cite{Schneider1986}. Firstly, we choose special starting point
$\vec{x}_{0}$ which has image positions $\vec{r}_{0}$ solved using an
analytical method. These starting points satisfy $x_{01}=0$ or
$x_{02}=0$ where $\vec{x}_{0}=(x_{01},x_{02})$. Secondly, let
$\vec{x}_{1}$ be a point near $\vec{x}_{0}$. The corresponding image
positions of $\vec{r}_{1}$ are given by the differential lens equation,
$d\vec{x}=Ad\vec{r}$, that is,
$\vec{r'}_{1}=\vec{r}_{0}+A^{-1}(\vec{r}_{0})(\vec{x}_{1}-\vec{x}_{0})$.
Thirdly, we insert $\vec{r'}_{1}$ into the lens Eq.
(\ref{equation6}) to find the corresponding source position
$\vec{x'}_{1}=\vec{x}(\vec{r'_1})$. The condition is
$|\vec{x}_{1}-\vec{x'}_{1}|<\varepsilon$. If this condition is not
satisfied, we take the next iterative step. For example, we use
$\vec{r_{1}'},\vec{x_{1}'}$ as starting points to do the next loop
of calculation. Another thing to note is that any steps of the
calculation should not cross a critical line, which is given by $\det
A=0$. In order to find all image positions of a given source, we
should choose suitable starting points in different areas separated
by critical lines.

With a specific configuration of lenses and source, we require two
conditions. Firstly, we should observe at least three images, that is,
three peaks in the FRB light curve. The flux ratio $I$ between every
two images should be smaller than a critical flux ratio $I_{f}$.
Secondly, there are at least three images which have the observed
time delay between every two images larger than a critical time
$t_{f}$. Given the position of source and the distance of source and
lenses, these conditions require the lens system to stay in some
specific areas. We define these specific areas as the cross section
$\sigma$. We use the Monte Carlo method to calculate the cross section.

\begin{figure}
  \centering
 \label{fig1}
\includegraphics[width=0.5\textwidth]{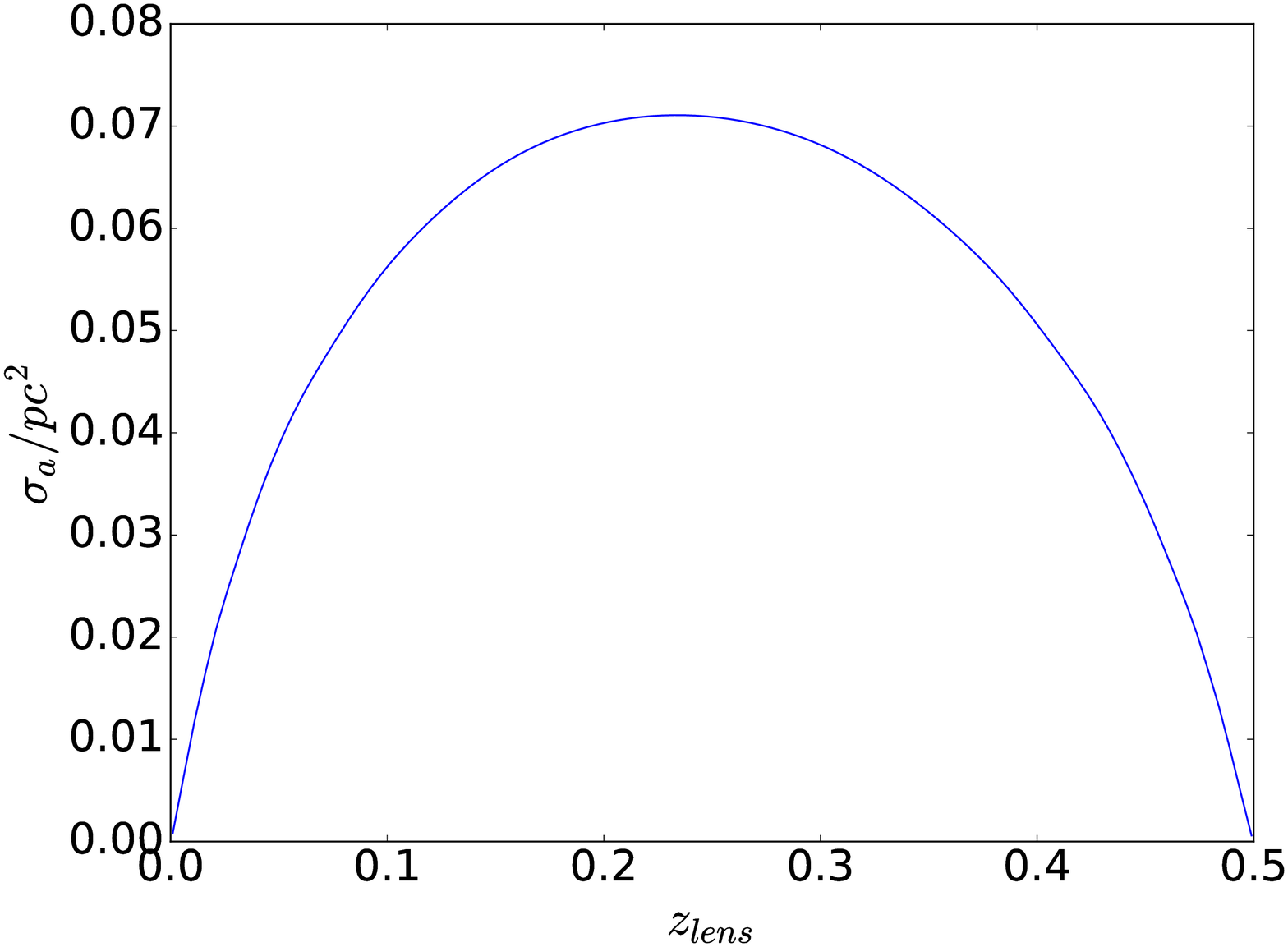}
\caption{The average cross section $\sigma_{a}$ of lenses at
different distances when the
  redshift of the source is $z=0.5$ and the fraction of dark matter in PBHs is
  $f=0.1$. The critical time $t_{f}=0.1$ ms and critical flux ratio $I_{f}=1000$ are used.}
\end{figure}

The separation of two lenses could affect the cross section. We
generate the PBH binaries and obtain the distribution of their
separations using the mechanism proposed by \cite{Nakamura1997} and
\cite{Sasaki2016}. Then, given the distance of source and lenses, we
calculate the average cross section by simply averaging the cross
section of a large number of PBH binaries staying in this distance.
Here, the average cross section
$\sigma_{a}(M,\mu_{1},\mu_{2},z_{s},z_{l})$ is a function of the
total mass of the lenses $M$, mass ratios $\mu_{1}$ and $\mu_{2}$,
redshift of source $z_{s}$ and lens system $z_{l}$. We calculate the average cross section of lenses at different distances
when the source is located at $z=0.5$ as an example. The result is shown in Fig.
1.

Next we calculate the lensing optical depth of a source,
which indicates the probability that this source is lensed
and the multi-peak structure is observable, at redshift $z_{s}$
from
\begin{equation}\label{equation11}
\tau=\int_{0}^{z_{s}}dX(z)(1+z)^{2}n_{L}\sigma_{a}(z).
\end{equation}
To obtain the estimated number of detected lensing events of FRBs by PBH
binaries, we need to convolve the optical depth for a single FRB
with the redshift distribution of FRBs. Similar to
\cite{Munoz2016} and \cite{Caleb2016}, we discuss two redshift
distributions of FRBs. The first one is a constant comoving number
density, that is, $\rho(z)=$constant. Observations show that the host
galaxy of FRB 121102 is a dwarf galaxy with low metallicity and
prominent emission lines \citep{Tendulkar2017}, sharing similar
properties with long gamma-ray bursts (GRBs). Meanwhile, long GRBs
trace the star formation history (SFH)
\citep{Wijers98,Wanderman10,Wang15}. If FRB 121102 is typical of the
wider FRB population, the rate of FRB would trace the SFH. So we
also consider that FRB number density is proportional to the SFH .
The SFH is \citep{Hopkins2006}
\begin{equation}\label{equation12}
\rho(z)=h\frac{a+bz}{1+\left(\frac{z}{c}\right)^d},
\end{equation}
with $a=0.0170$, $b=0.13$, $c=3.3$, and $d=5.3$ and
$h=0.7$\citep{Cole2001,Wang2013} is used. The comoving volume of a
shell of width $dz$ at redshift $z$ is $dV(z)=[4\pi \chi^2
(z)/H(z)]dz $. Introducing the Gaussian cutoff to represent an
instrumental signal-to-noise (S/N) threshold, we get the distribution of
FRBs;
\begin{equation}\label{equation13}
N_{FRB}(z)=\mathbb{N}_{FRB}\frac{\rho(z)\chi(z)^{2}}{H(z)(1+z)}e^{-d_{L}^{2}/[2d_L^{2}(z_{cut})]},
\end{equation}
where $\mathbb{N}_{FRB}$ is the normalizing factor and the term
$(1+z)$ accounts for the effect of cosmological time dilation.
We choose $\Omega_{vac}=0.714$, $\Omega_M=0.286$, and
$H_0=69.6$ km s$^{-1}$Mpc$^{-1}$ in our calculation.

Then we can calculate the integrated optical
depth $\bar{\tau}$
\begin{equation}\label{equation14}
\bar{\tau}=\int dz\tau(z)N_{FRB}(z).
\end{equation}
Inter-channel dispersion will broaden the FRB pulse width. Meanwhile,
its intrinsic pulse profile and scattering with the intergalactic
medium will also contribute the total pulse width. Similar to
\cite{Munoz2016}, we choose 1 ms and 0.1 ms as the critical times
$t_{f}$. For the critical flux ratio $I_{f}$, we adopt three
different values, that is, $10^3$, 100, and 10. The cutoff
redshift is chosen as $z_{cut}=0.5$ \citep{Munoz2016}, which fits
the current FRB catalog well for both two distributions of FRBs. In
Table \ref{table1}, we show the integrated optical depth
$\bar{\tau}$ for different fractions of dark matter
$f=10^{-1},10^{-2},10^{-3}$ in PBHs for FRB number density
proportional to SFH. In Table \ref{table2}, we show $\bar{\tau}$ for
different $f=10^{-1},10^{-2},10^{-3}$ in PBHs for a constant FRB
number density.

\begin{table*}
\centering \caption{The integrated optical depth $\bar{\tau}$ for
different parameters for a constant FRB comoving number density.\label{table2}}
\begin{tabular}{|c|c|c|c|c|c|c|c|}
\hline
$t_{f},I_{f}$ &$10^{-4}s,10^{3}$&$10^{-3}s,10^{3}$&$10^{-4}s,10^{2}$&$10^{-3}s,10^{2}$&$10^{-4}s,10^{1}$&$10^{-3}s,10^{1}$\\
\hline
$f=10^{-3}$& $1.30\times10^{-4}$ & $7.92\times10^{-5}$ & $1.98\times10^{-5}$ & $8.85\times10^{-6}$& $1.02\times10^{-6}$ & $4.03\times10^{-7}$\\
\hline
$f=10^{-2}$& $4.59\times10^{-3}$ & $2.67\times10^{-3}$ & $6.79\times10^{-4}$ & $2.89\times10^{-4}$ & $3.59\times10^{-5}$ & $1.41\times10^{-5}$ \\
\hline
$f=10^{-1}$& $4.76\times10^{-2}$ & $2.77\times10^{-2}$ & $7.02\times10^{-3}$ & $3.02\times10^{-3}$ &$3.80\times10^{-4}$ &$1.46\times10^{-4}$ \\
\hline
\end{tabular}
\end{table*}

We  choose the above values for $t_f$ and
$I_f$ because the duration time of FRBs is on the order of 1 ms
\footnote{http://www.astronomy.swin.edu.au/pulsar/frbcat}. Therefore, the
choice of critical time $t_{f}=1$ ms is reasonable. We
compare the repeating FRBs with the lensing FRBs since repeating
FRBs also have multi-peak structures which might mix with the
lensing FRBs. It turns out that the time interval between
successive bursts for the repeating FRB 121102 is larger than
several tens of seconds \citep{Wang2017}. Therefore, the lensing
FRBs can be discriminated from typical repeating FRBs. To date,
multiple bursts have only been observed for the repeating FRB 121102. Fluxes are 670 mJy \citep{Chatterjee2017} and 20 mJy
\citep{Spitler2016} for the brightest and dimmest burst,
respectively, giving a flux ratio between them of about 33. Therefore,
the choice of the critical flux ratio $I_{f}$ is reasonable.

Since the integrated optical depth $\bar{\tau}$ is very small, the
expected number of lensing FRBs by PBH binaries is
\begin{equation}\label{equation15}
N_{exp}=N\times P=N\times (1-e^{-\tau})=N\times \tau.
\end{equation}
\cite{Connor2016} estimated that the Canadian hydrogen
intensity mapping experiment (CHIME) could detect about $10^4$ FRBs
per year. We could expect the detection of FRBs strongly lensed by
PBH binaries. For example, the number of strongly lensed FRBs is
about 36 for $t_{f}=1$ ms, $I_{f}=100$ and $f=0.1$ per year.
Because most of the FRBs are below $z=1.5$ in our calculation when choosing $z_{cut}=0.5$, the band of CHIME is not severely affected by redshift.

\section{Numerical simulations}\label{sec:simulation}

We further simulate the gravitational lensing events of FRBs by PBH
binaries to verify the analytical calculation in the above section. Assuming the distribution
of PBH binaries is isotropic and uniform in space, we generate
the PBH binaries following \cite{Sasaki2016}. As mentioned in the
previous section, we consider the distribution of FRBs following the
SFH \citep{Caleb2016}. For the constant comoving number
density, the difference of the simulation resulting from the SFH case
is less than $30\%$.

\begin{table*}
\centering \caption{The results of the simulation of $2\times
10^{5}$ FRBs for different parameters.\label{table3}}
\begin{tabular}{|c|c|c|c|c|c|c|}
\hline
\multicolumn{7}{|c|}{Choosing the smallest $b/\Sigma_{cr}^{-1}$}\\
\hline
$t_{f},I_{f}$   & $10^{-3}s,10^{1}$ & $10^{-4}s,10^{1}$ & $10^{-3}s,10^{2}$ & $10^{-4}s,10^{2}$ & $10^{-3}s,10^{3}$ & $10^{-4}s,10^{3}$\\
\hline
$f=10^{-3}$& 0 & 0 & 0 & 0 & 3 & 7 \\
\hline
$f=10^{-2}$& 0 & 1 & 7 & 34 & 121 & 257\\
\hline
$f=10^{-1}$& 0 & 9 & 82 & 275 & 1085 & 2256\\
\hline
\multicolumn{7}{|c|}{Choosing the smallest $b/\sqrt{\Sigma_{cr}^{-1}}$}\\
\hline
$t_{f},I_{f}$   & $10^{-3}s,10^{1}$ & $10^{-4}s,10^{1}$ & $10^{-3}s,10^{2}$ & $10^{-4}s,10^{2}$ & $10^{-3}s,10^{3}$ & $10^{-4}s,10^{3}$\\
\hline
$f=10^{-3}$& 0 & 0 & 3 & 5 & 5 & 8 \\
\hline
$f=10^{-2}$& 0 & 2 & 16 & 38 & 125 & 257\\
\hline
$f=10^{-1}$& 0 & 6 & 98 & 296 & 1079 & 2281\\
\hline
\multicolumn{7}{|c|}{Choosing the smallest $\Delta \theta/\sqrt{\Sigma_{cr}^{-1}}$}\\
\hline
$t_{f},I_{f}$   & $10^{-3}s,10^{1}$ & $10^{-4}s,10^{1}$ & $10^{-3}s,10^{2}$ & $10^{-4}s,10^{2}$ & $10^{-3}s,10^{3}$ & $10^{-4}s,10^{3}$\\
\hline
$f=10^{-3}$& 0 & 0 & 0 & 0 & 3 & 6 \\
\hline
$f=10^{-2}$& 0 & 0 & 14 & 34 & 115 & 241\\
\hline
$f=10^{-1}$& 0 & 8 & 108 & 284 & 1023 & 2129\\
\hline
\end{tabular}
\end{table*}

\begin{figure}

  \centering

\includegraphics[width=0.5\textwidth]{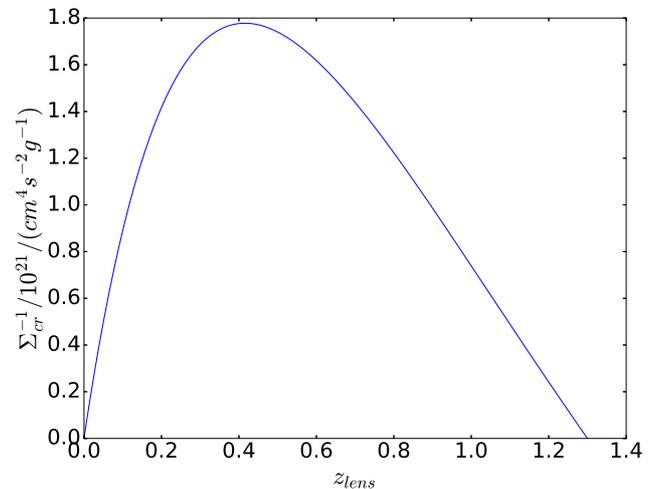}
\caption{The lensing kernel $\Sigma_{cr}^{-1}$ of lenses at
different redshifts when the
  redshift of the source is $z=1.3$. \label{kernel}}
\end{figure}

\begin{figure}
  \centering
  \subfigure{
  \label{Fig2.sub.1}
  \includegraphics[width=0.5\textwidth]{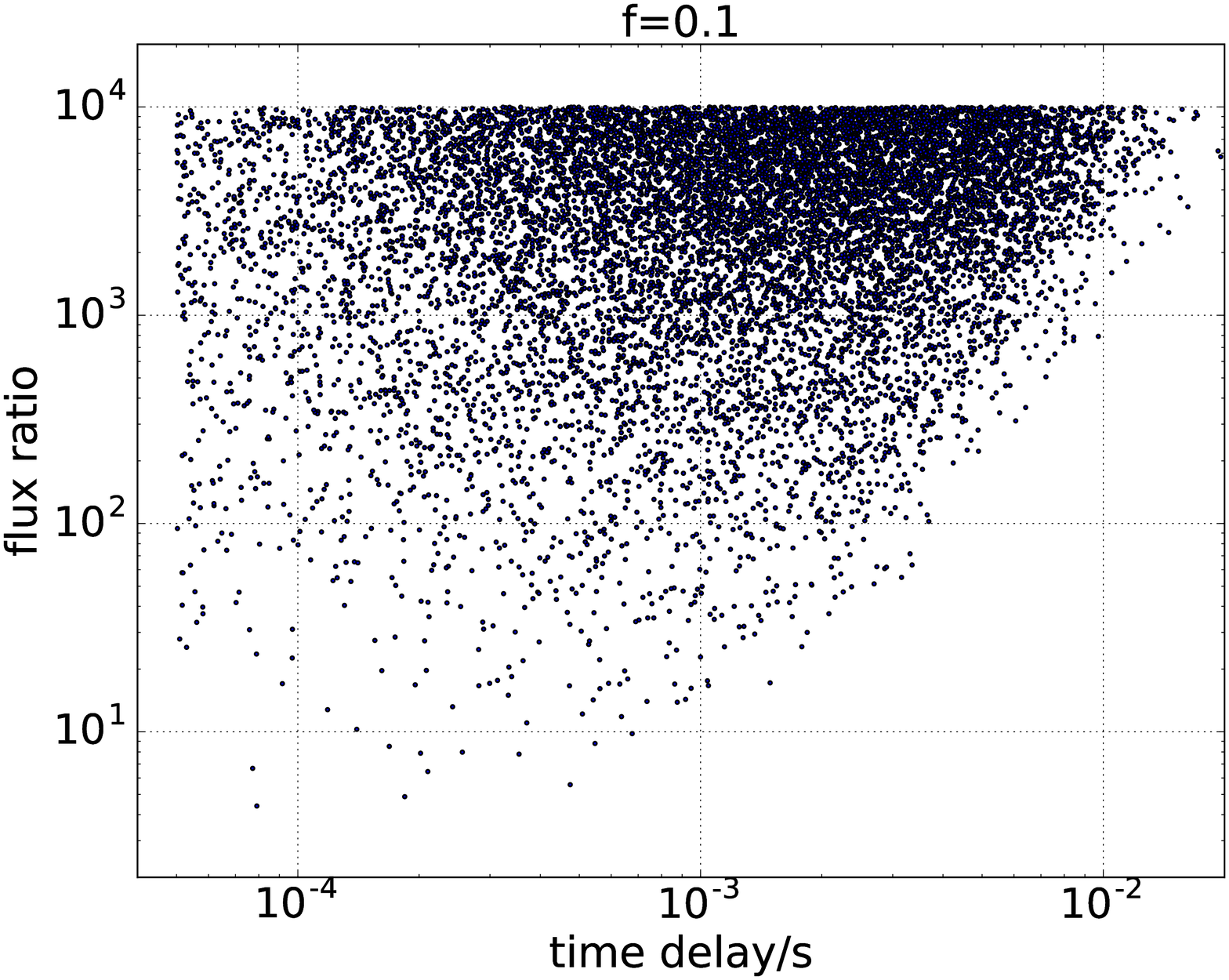}}
  \subfigure{
  \label{Fig2.sub.2}
  \includegraphics[width=0.5\textwidth]{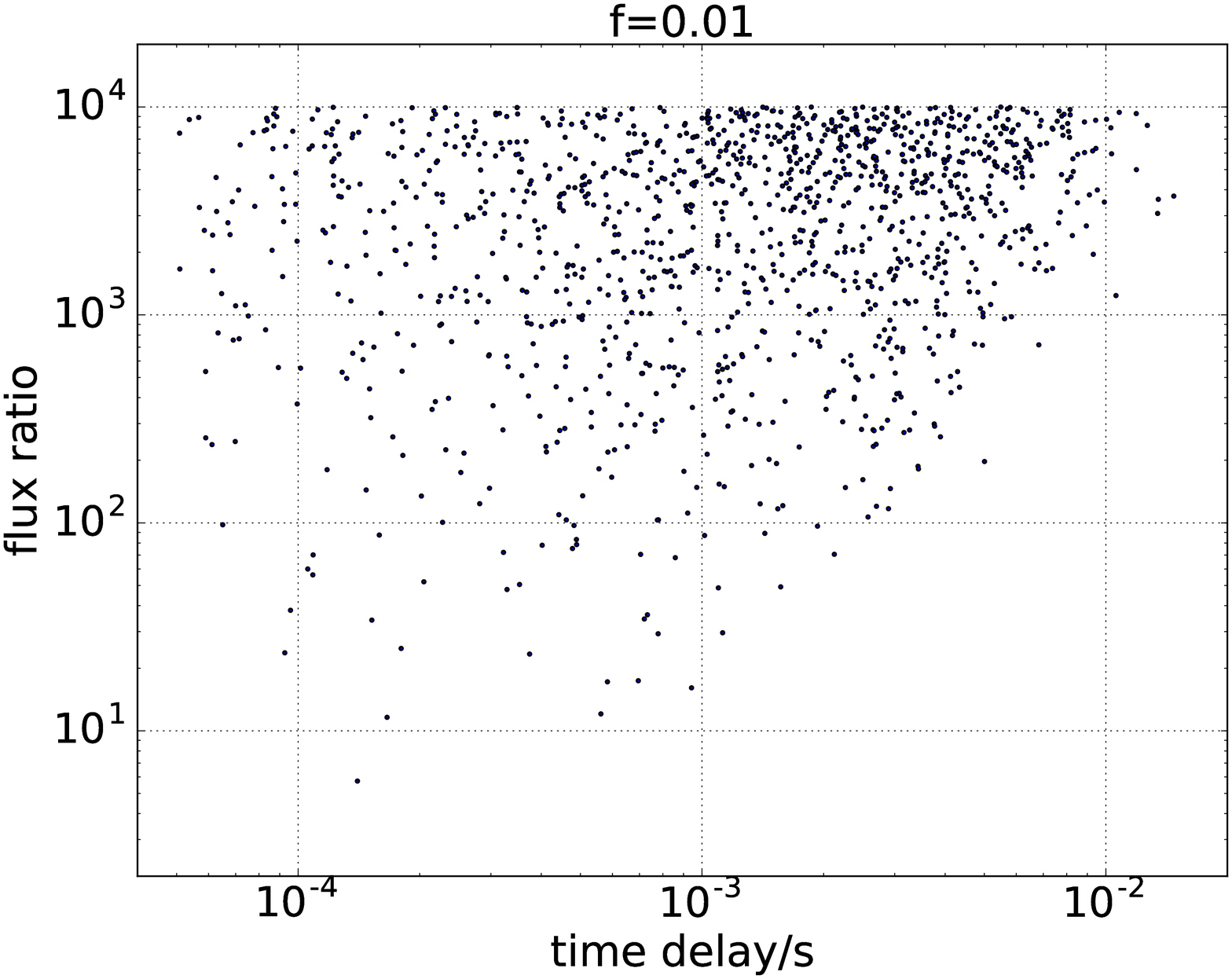}}
  \subfigure{
  \label{Fig2.sub.3}
  \includegraphics[width=0.5\textwidth]{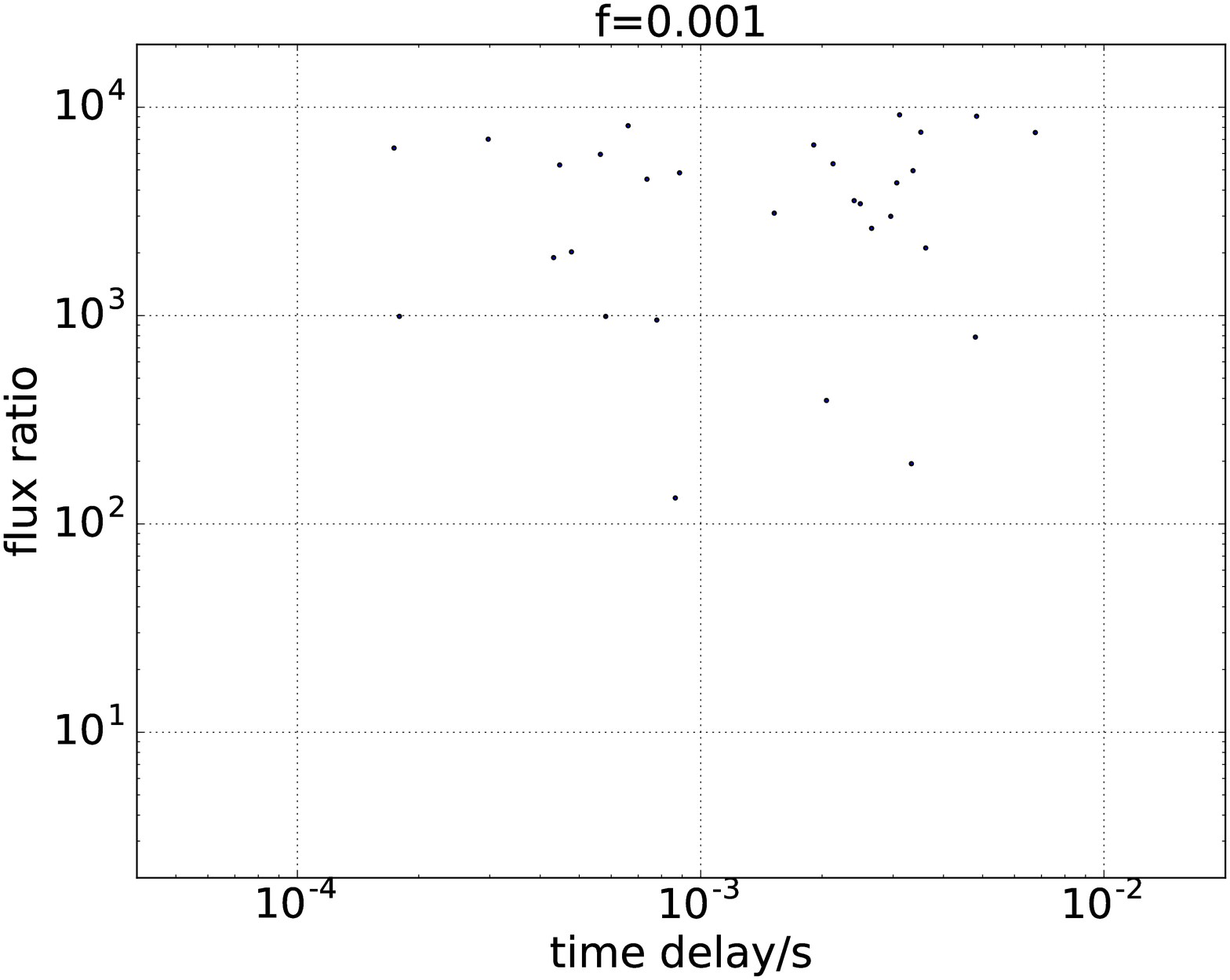}}
  \\
  \caption{Simulations of $2\times 10^{5}$ FRBs with different fractions of dark matter $f=10^{-1},10^{-2},10^{-3}$, respectively. All three panels show all the lensing events with flux ratio $I<10^{4}$ and time delay $t>0.5\times 10^{-4}$ s.}\label{fig2}
\end{figure}

We simulate $2\times 10^{5}$ FRBs for different fractions of dark
matter $f=10^{-1},10^{-2},10^{-3}$ in PBHs, respectively. We choose
$\Omega_{vac}=0.714$, $\Omega_M=0.286$, and $H_0=69.6\rm km~s^{-1}Mpc^{-1}$
in our calculation. For each FRB, we consider the PBH binaries which
could provide the strongest lensing effect. The lensing efficiency
is indicated by lensing kernel
\begin{equation}\label{equation16}
\Sigma_{cr}^{-1}=\frac{4\pi GD_{ds}D_{d}}{D_{s}},
\end{equation}
which is proportional to $D_{ds}D_{d}/D{s}$
\citep{Sheldon2002,Courbin2002}. It describes the geometry of the
lense-source system. A larger $\Sigma_{cr}^{-1}$ indicates a stronger
lensing effect. As an example, the lensing kernel of the lenses at
different redshifts when the source is at redshift $z=1.3$ is shown
in Fig. \ref{kernel}.  The distance of PBH binaries from the line
of sight can also affect the lensing effect. Because the
lensing events depend on other parameters, such as the orbital
elements of the binaries, the strongest lensing event might not give the
expected observable multi-peak FRBs. Meawhile, the magnification
and time delay are also required. Therefore, we use several criteria to show
the reliability of our results. We choose the PBH binaries by
finding the smallest value of $\Delta
\theta/\sqrt{\Sigma_{cr}^{-1}}$, $b/\sqrt{\Sigma_{cr}^{-1}}$ or
$b/\Sigma_{cr}^{-1}$, where $\theta$ is the angular distance of the
PBH binary from the line of sight, and $b$ is the linear distance of the
PBH binary from the line of sight, that is, the impact parameter. We
calculate the positions of images, time delay, and flux ratio of
every second image.   Then, we check whether they could conform with
the conditions concerning time delay and flux ratio. We record all the
lensing events with flux ratio $I<10^{4}$ and time delay
$t>0.5\times 10^{-4}$ s and give the results of the
simulations in Table \ref{table3} for all three different ways of
choosing PBH binaries. We find that the difference due to this latter choice is not
obvious. Therefore, the choice of criteria cannot significantly affect the results. They all provide a lower limit of
observable multi-peak FRBs. We further show the results for
choosing PBH binaries by finding the smallest value of
$b/\Sigma_{cr}^{-1}$ in Fig. \ref{fig2}. The number of lensing
events is strongly influenced by the fraction of dark matter, $f$,
critical time delay, $t_{f,}$ and critical flux ratio, $I_{f}$.

There are some differences between the analytical results in Sect.
\ref{sec:Optical Depth} and numerical simulations in Sect.
\ref{sec:simulation}. These differences are reasonable because we make some assumptions in the
calculation in Sect.
\ref{sec:Optical Depth}. For example, the average cross section used to calculate the optical
depth is not very accurate. In Sect. 3, we use the Monte Carlo method to calculate
the cross section since the allowed regions are not regular, as only a
single lense is used.

\section{Discussion and Conclusion}\label{sec:conclusion}
FRBs are promising tools to probe MACHOs. Firstly, FRBs are
strong and short events, which means that they could serve as good candidates to
probe the lensing events. As a result of lensing by black hole
binaries, we should detect multi-peak FRBs. Since FRBs are strong
and short, their multi-peak structure might be observable. Secondly,
the rate of FRBs can be as high as a few thousand per day for the
whole sky. Lastly, in the upcoming or ongoing surveys, the
observation of a large number of FRBs is promising.

In this paper, we estimate the possibility of detecting lensing
events of FRBs by PBH binaries. The typical time interval between
different images is roughly 1 ms. The flux ratio between different
images is from approximately 10 to $10^3$. With an expected detection
rate of FRBs by the upcoming experiment CHIME of $10^4$ FRBs per
year, we could expect five detections of multi-peak FRBs with time
intervals larger than 1 ms and flux ratios less than $10^3$ per year
in the upcoming FRB surveys if the fraction of dark matter in MACHOs
is $f=0.01$. If the lensing events are detected, we could analyze
the structure of the peaks of FRBs to infer the separation and  masses of
two lenses to verify the primordial black hole scenario for
gravitational-wave events. Alternatively, if no FRBs are lensed, the
strongest constraints will be those put on the fraction of dark matter in the
form of compact objects.

In our calculation, we use the formation mechanism of PBH binaries
proposed in \cite{Nakamura1997}.  Tidal force from the nearby PBH
provides the angular momentum of the PBH binaries. \cite{Bird2016} proposed that PBH binaries are formed due to
the gravitational radiation when they pass by one another. We wish to
point out that in our calculation, these two mechanisms do not have
obvious differences. In the first mechanism, most of the PBH binaries
will have an eccentricity very close to one. The movement of PBHs
in the binary system could be very similar to movement when they are
free. If we assume the same number density of PBHs, the results from
two mechanisms would be very similar.

\cite{Cordes2017} claim that plasma lenses in FRB
host galaxies can produce multiple bursts with different apparent
strengths, arrival times, and dispersion measurements. There are
several differences between our results and theirs. Firstly, the
positions and "gain" (or amplification) of burst images are
different for different frequencies in Cordes et al. (2017).
However, in our discussion, the lensing effect caused by PBH
binaries would be the same for any frequency. Secondly, the DMs of
different images of bursts are different in their calculations. As
for images caused by PBH lensing, the DMs of burst images
contributed by the host galaxy are the same. Thirdly, their theory
allows for the existence of larger time intervals between images,
which could interpret the repeating FRB 121102 qualitatively. In our
paper, we expect to distinguish between repeating FRBs and multiply
imaged bursts caused by PBH binaries.

\section*{Acknowledgements}
We thank the anonymous referee for constructive suggestions that have allowed
us to improve our manuscript. This work is supported by the National
Basic Research Program of China (973 Program, grant No.
2014CB845800) and the National Natural Science Foundation of China
(grants 11422325 and 11373022), the Excellent Youth Foundation of
Jiangsu Province (BK20140016).

\newpage

\end{document}